%% file: arxiv_v1.tex
\documentclass[12pt]{article}
\usepackage{epsfig}
\usepackage{graphicx}
\usepackage{bm}

\input econfmacros-csqcd4.tex
\def\Title#1{\begin{center} {\Large {\bf #1} } \end{center}}

\begin{document}

\Title{Overview of Crystalline Color Superconductors }

\bigskip\bigskip


\begin{raggedright}

{\it 
Massimo Mannarelli$^{1}$\\
\bigskip
$^{1}$INFN, Laboratori Nazionali del Gran Sasso, Via G. Acitelli, 22, I-67100 Assergi (AQ), Italy\\
\bigskip
}
\end{raggedright}

\section{Introduction}
Inhomogeneous phases may appear when a stress is applied to a system and the system can minimize the free energy breaking the rotational invariance. Various examples are known in Nature of this sort, as the paramagnetic to ferromagnetic  phase transition, or the fluid/solid phase transition. If the rotational symmetry is broken down to a discrete symmetry, the system is typically named a  crystal. Crystals generally can form in two different ways, by the presence of an attractive interaction between the elementary constituents or by compression induced by an external agent. Standard crystals, like metallic crystal,  form by means of the first mechanism when the Coulomb interaction between ions becomes stronger than the thermal energy. An example of the  second mechanism is solid helium. In this case the short range repulsive force between ultracold helium atoms prevents the collapse of the system under an external pressure, inducing the formation of an ordered phase.

\section{Inhomegeneous quark matter}
Whether one of the two mechanisms described above is at work for quark matter is not obvious. Perturbatively, it is known that the strong interaction between quarks has both an attractive and a repulsive channel, depending on the color degrees of freedom. Thus, both mechanisms can in principle work in quark matter for producing a solid phase.

In the first place,  for having deconfined  quark matter it is necessary a large energy density, which can be realized in heavy ion collisions (HICs) or in compact stellar objects (CSOs),  corresponding to stars having a radius of about $10$ km and a mass of about a solar mass. The typical reference energy in strong interactions is $\Lambda_{QCD} \sim 200$ MeV.
In  HICs the baryonic density is small and the temperature is larger than $\Lambda_{QCD}$, meaning that the formation of a solid-like phase is improbable. On the other hand, CSOs are relatively cold  stars with a temperature much less  than $\Lambda_{QCD}$ and a quark chemical potential that can reach $400-500$ MeV.  Thus,  CSOs are cold and dense, an ideal environment for the formation of a solid-like phase. 

It is unclear whether quark mater is present in CSOs, but if it is present it is likely that the attractive color interaction induces the transition to a color superconductor (CSC). In particular, at asymptotic density quark matter is expected to become a  CSC in the color-flavor locked (CFL) phase~\cite{Alford:1998mk}.  The reason of color superconductivity is that the  estimated critical temperature of CSCs is at least of the order of few MeV,  much larger than the  tens of KeV temperature of  few seconds old CSOs. The reason of CFL phase is that in this phase $u, d, s$ quarks of all colors pair coherently maximizing the free-energy gain. The CFL phase  is extremely  robust, corresponding to the  pairing pattern between all quarks that preserve the largest global  symmetry group.

\begin{figure}[htb]
\begin{center}
\includegraphics[width=8cm]{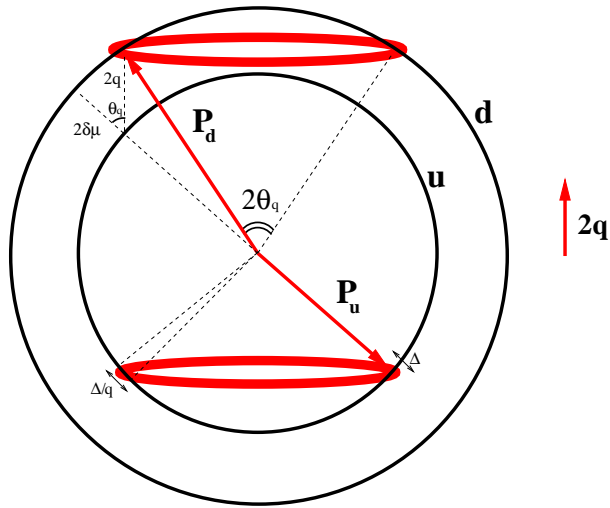}
\end{center}
\caption{\label{Fig:Pairing} (color online). Pictorial description of the CCSC pairing regions for a single plane wave structure in two-flavor quark matter. The quark pairs have momentum   $\bm P_u + \bm P_d = 2 \bm q$.   The pairing regions correspond to the  two red ribbons  on the top  the Fermi spheres and are characterized by an opening angle $2 \vartheta_q \simeq  67^{\circ}$, thickness  $\sim \Delta$ and  angular width  $\sim \Delta/q$.}
\end{figure}

The CFL phase is homogeneous and robust, thus it is unlikely to become a solid, unless a strong stress acts on it. There are  two different kinds of stress that can lead to the crystallization. The surface tension of quark matter can lead to the formation of a solid phase close to the surface of strange stars. In this case, clusters of  strange matter known as strangelets arrange themselves in a rigid structure. The realization of this  crust depends on the value of the surface tension between strange quark matter and the vacuum~\cite{Jaikumar:2005ne}. A second possibility is that  the CSC gap parameter is spatially modulated, forming a crystalline color superconductor (CCSC). In this case, quark matter would form a solid in bulk. The stress responsible for the formation of the CCSC structure relies on the mismatch of the Fermi spheres produced by the combined effect of the electroweak equilibrium and the strange quark mass, $M_s$.  The resulting mismatch between the Fermi spheres is proportional to $M_s^2/\mu$, where 
$\mu$ is the average  quark chemical potential.  For sufficiently high mismatches 
the system can minimize the free energy by restricting the  pairing to selected quark flavors, as in the two-flavor color superconducting phase, or by quark pairing in restricted regions of the Fermi spheres, as in the  CCSC phase~\cite{Alford:2000ze, Rajagopal:2006ig,Anglani:2013gfu}. In particular,  the restriction of pairing to certain regions of the Fermi sphere results in quark pairs  having  total momentum, $2 \bm q$, as illustrated in Fig.~\ref{Fig:Pairing} for the two-flavor case with one couple of pairing regions.

Formally,  the general three-flavor CCSC condensate  is given by
\begin{equation}
\small{\langle
0|\psi_{iL}^\alpha\psi_{jL}^\beta|0\rangle = {-\langle
0|\psi_{iR}^\alpha\psi_{jR}^\beta|0\rangle} \propto
 \sum_{I=1}^3\
 \Delta_I \varepsilon^{\alpha\beta\gamma}\epsilon_{ijk} \hspace{-.3cm}
\sum_{\bm q_{I}^{m}\in\{\bm q_{I}\}} e^{2i\bm q_{I}^{m}\cdot\bm x}}\,,
 \label{condensate-crystal}
\end{equation}
where $\psi_{jL(R)}^\alpha$ are left (right) handed fermionc fields with flavor $i$ and color $\alpha$ and    $\varepsilon^{\alpha\beta\gamma}$ and $\epsilon_{ijk}$ are the completely antisymmetric Levi-Civita symbols in 
color and flavor space, respectively.  The color structure is determined by the requirement of interaction in the antitriplet channel and the flavor structure is determined by the requirement of $s-$wave interaction. The condensate with index $I$ corresponds to pairing between quarks whose flavor and color is not $I$. The modulation of the $I$'th
condensate is defined by the vectors $\bm q_{I}^{m}$, where $m$ is the index which identifies the elements of the  set  $\{\bm q_{I}\}$.  In position space, this corresponds to condensates that vary like $\sum_m \exp(2 i {\bm q_I^m}\cdot {\bm x})$, meaning that
the ${\bm q}_I^m$'s are the reciprocal vectors which define the crystal structure of the condensate.  The CCSC phase is  the QCD analogue of a form of non-BCS pairing
first proposed by Larkin, Ovchinnikov, Fulde and Ferrell
(LOFF) \cite{LO, FF}  for systems with Fermi mismatches exceeding the  Chandrasekhar-Clogston  limit \cite{Chandrasekhar, Clogston}).

Distinguishing the phenomenological signature of  standard nuclear matter and deconfined quark matter is hard, because we have poor knowledge of both the properties of nuclear matter and of quark matter at densities above the nuclear saturation density. In particular, the poorly constrained  equation of state of quark mater can easily reproduce the properties of the poorly known equation of state of nuclear matter above saturation density~\cite{Alford:2004pf}. In~\cite{Mannarelli:2014ija} and  \cite{Mannarelli:2015jia}  we have investigated  the possible signature of torsional oscillations of the crust of strange stars. These oscillations are particularly interesting because the restoring force  is the elastic shear stress and the shear modulus of the CCSC phase is much larger than the shear modulus of the neutron star crust. The typical frequency obtained are of the order of the kHz. In particular in \cite{Mannarelli:2015jia} a nonbare strange star model having a CCSC crust surmounted by a standard ionic crust was considered.  We found that  even if a small fraction  of the energy of a Vela-like glitch is conveyed to a torsional oscillation, the ionic crust will likely break. The reason is that the very rigid and heavy CCSC crust layer will absorb only a small fraction of the oscillation energy, leading to a large torsional oscillation of the ionic crust eventually exceeding the breaking strain of nuclear matter.

\subsection*{Acknowledgement}

We express our thanks to the organizers of the CSQCD IV conference for providing an 
excellent atmosphere which was the basis for inspiring discussions with all participants.
We have greatly benefitted from this.


\end{document}

%% file: econfmacros-csqcd4.tex
\def\beq{\begin{equation}}
\def\eeq#1{\label{#1}\end{equation}}
\def\eeqn{\end{equation}}

\def\beqa{\begin{eqnarray}}
\def\eeqa#1{\label{#1}\end{eqnarray}}
\def\eeqan{\end{eqnarray}}

\let\bar=\overbar

\def\Dslash{\not{\hbox{\kern-4pt $D$}}}
\def\dslash{\not{\hbox{\kern-2pt $\del$}}}

\def\msb{{\bar{\ssstyle M \kern -1pt S}}}

\usepackage{fancyhdr,graphicx}

%% file: arxiv_v1.bbl
\begin{thebibliography}{10}

\bibitem{Alford:2000ze}
M.~G. Alford, J.~A. Bowers, and K.~Rajagopal.
\newblock {Crystalline color superconductivity}.
\newblock {\em Phys.Rev.}, D63:074016, 2001.

\bibitem{Alford:2004pf}
M.~G. Alford, M.~Braby, M.~W. Paris, and S.~Reddy.
\newblock {Hybrid stars that masquerade as neutron stars}.
\newblock {\em Astrophys.J.}, 629:969--978, 2005.

\bibitem{Alford:1998mk}
M.~G. Alford, K.~Rajagopal, and F.~Wilczek.
\newblock {Color flavor locking and chiral symmetry breaking in high density
  QCD}.
\newblock {\em Nucl.Phys.}, B537:443--458, 1999.

\bibitem{Anglani:2013gfu}
R.~Anglani, R.~Casalbuoni, M.~Ciminale, N.~Ippolito, R.~Gatto, et~al.
\newblock {Crystalline color superconductors}.
\newblock {\em Rev.Mod.Phys.}, 86:509--561, 2014.

\bibitem{Chandrasekhar}
B.~S. Chandrasekhar.
\newblock {a Note on the Maximum Critical Field of High-Field Superconductors}.
\newblock {\em Appl. Phys. Lett.}, 1:7--8, 1962.

\bibitem{Clogston}
A.~M. Clogston.
\newblock Upper limit for the critical field in hard superconductors.
\newblock {\em Phys. Rev. Lett.}, 9:266--267, 1962.

\bibitem{FF}
P.~Fulde and R.~A. Ferrell.
\newblock Superconductivity in a strong spin-exchange field.
\newblock {\em Phys. Rev.}, 135:A550--A563, 1964.

\bibitem{Jaikumar:2005ne}
P.~Jaikumar, S.~Reddy, and A.~W. Steiner.
\newblock {The Strange star surface: A Crust with nuggets}.
\newblock {\em Phys.Rev.Lett.}, 96:041101, 2006.

\bibitem{LO}
A.~I. Larkin and Y.~N. Ovchinnikov.
\newblock Nonuniform state of superconductors.
\newblock {\em Zh. Eksp. Teor. Fiz.}, 47(3):1136--1146, 1964.

\bibitem{Mannarelli:2014ija}
M.~Mannarelli, G.~Pagliaroli, A.~Parisi, and L.~Pilo.
\newblock {Electromagnetic signals from bare strange stars}.
\newblock {\em Phys.Rev.}, D89:103014, 2014.

\bibitem{Mannarelli:2015jia}
M.~Mannarelli, G.~Pagliaroli, A.~Parisi, L.~Pilo, and F.~Tonelli.
\newblock {Torsional oscillations of nonbare strange stars}.
\newblock 2015.

\bibitem{Rajagopal:2006ig}
K.~Rajagopal and R.~Sharma.
\newblock {The Crystallography of Three-Flavor Quark Matter}.
\newblock {\em Phys.Rev.}, D74:094019, 2006.

\end{thebibliography}
